\documentclass[12pt]{article}
\usepackage[dvips]{epsfig}
\newcommand{\CO}{{CompHEP}}
\newcommand{\noi}{\noindent}
\newcommand{\eeinto}   {$e^+e^- \longrightarrow \:$}
\newcommand{\geinto}   {$\gamma e \longrightarrow \:$}
\newcommand{\into}   {$\longrightarrow \:$}
\newcommand{\entb}{$e \nu t b \: $}
\newcommand{\gentb}{$\gamma e \longrightarrow  \nu \bar{t} b \: $}
\newcommand{\hww}{$HWW \: $}
\newcommand{\nbbw}{$\nu \,b\, \bar{b}\, W\: $}

\newcommand{\mt}{$m_{t} \: $}
\newcommand{\mh}{$M_{H} \: $}
\newcommand{\ee}{$e^+e^-\: $}
\newcommand{\vtb}{$|V_{tb}| \: $}

\newcommand{\bb}{$b\,\bar{b}\: $}

\newcommand{\qq}{$q\,\bar{q}\: $}

\newcommand{\hnull}{$H^0 \: $}

\newcommand{\SQRTSGE}{$\sqrt{s_{\gamma e}}\:$}
\newcommand{\SQRTSEE}{$\sqrt{s_{e^+e^-}}\:$}
\newcommand{\less}{\stackrel{ <}{\sim}}

\newcommand{\Ptb}{$p_{\perp}^b\:$}

\newcommand{\Pt}{$p_\perp\:$}
\newcommand{\Ptt}{$p_{\perp}^t\:$}

\def\PL #1 #2 #3 {Phys. Lett. {\bf#1}              (#3)  #2}
\def\MPL #1 #2 #3 {Mod. Phys. Lett. {\bf#1}        (#3)  #2}
\def\IMP #1 #2 #3 {Int. Mod. Phys.  {\bf#1}        (#3)  #2}
\def\NP #1 #2 #3 {Nucl. Phys. {\bf#1}              (#3)  #2}
\def\PR #1 #2 #3 {Phys. Rev. {\bf#1}               (#3)  #2}
\def\PP #1 #2 #3 {Phys. Rep. {\bf#1}               (#3)  #2}
\def\PRL #1 #2 #3 {Phys. Rev. Lett. {\bf#1}        (#3)  #2}
\def\CPC #1 #2 #3 {Comp. Phys. Commun. {\bf#1}     (#3)  #2}
\def\ANN #1 #2 #3 {Annals of Phys. {\bf#1}         (#3)  #2}
\def\APP #1 #2 #3 {Acta Phys. Pol. {\bf#1}         (#3)  #2}
\def\ZP  #1 #2 #3 {Z. Phys. {\bf#1}                (#3)  #2}
\def\NIM  #1 #2 #3 {Nucl. Instr. and Meth. {\bf#1} (#3)  #2}
 
\newcommand{\BS}{\bigskip}

\newcommand{\lone}{$F_1^L\: $}
\newcommand{\ltwo}{$F_2^L\: $}
\newcommand{\rone}{$F_1^R\: $}
\newcommand{\rtwo}{$F_2^R\: $}

\begin{document}

\pagestyle{empty}
\section*{
\vspace{4cm}
\begin{center}
\LARGE{\bf
 Single Top and Light Higgs    
 at TeV Energy $\gamma e$ Colliders
       }\\
\end{center}
}

\vspace{2.5cm}
\large
\begin{center}
E. Boos$^{1,2}$, A. Pukhov$^2$, M. Sachwitz$^3$ and H. J. Schreiber$^3$ \\ 
\bigskip \bigskip
$^1$Institut f\"{u}r Kernphysik, TU, Darmstadt, FRG \\  
$^2$Institute of Nuclear Physics, Moscow State University, 119899,
Moscow, Russia \\
$^3$DESY-Institut f\"{u}r Hochenergiephysik, Zeuthen, FRG \\
\end{center}

\vspace{7.0cm}
\noindent Contribution to the Proceedings of the 'ECFA/DESY Study on Physics 
and Detectors for the Linear Collider', DESY 97-123E, ed. by R.Settles.
\newpage
\pagestyle{plain}
\pagenumbering{arabic}
%----------------------------------------------------------------------
%----------------------------------------------------------------------

%----------------------------------------------------------------------
% ABSTRACT
%----------------------------------------------------------------------
%----------------------------------------------------------------------
\begin{center}
\section*{Abstract}
\end{center}
%\abstract{
Results of complete tree level calculation of the single top and light Higgs
production in the reaction \geinto \nbbw at the Next Linear Collider are 
presented. In addition, the contributions of anomalous operators to the 
$Wtb$ and $WWH$ couplings are included into the complete 4-body consideration.
The sensitivity for probing the structure of the couplings 
in a model independent way is analyzed.
%}

\BS\BS\BS

\large
%----------------------------------------------------------------------
% INTRODUCTION
%----------------------------------------------------------------------
\section{Introduction}

The top quark, by far the heaviest established elementary particle, 
is not only a further confirmation of the Standard Model (SM), 
but it also poses new questions.
One example is the spectacular numerical coincidence between the vacuum
expectation value $v/\sqrt{2}$ = 175 GeV
and the top quark mass, measured by the CDF and D0 collaborations
\cite{cdfd0} to be 175$^{+6}_{-6}$ GeV, and extracted indirectly from fits of
precision electroweak LEP data as 177$^{+7+16}_{-7-19}$ GeV \cite{ew}.
It is an open question whether or not this is due to 
fundamental physics relations or is only accidental.
The heavy top quark decays electro-weakly before hadronization 
\cite{bigi} and therefore it could provide a first window to 
help understand the
nature of the electroweak symmetry breaking \cite{peccei}.
In this context, reactions involving  a light Higgs 
boson and top quark production as intermediate states are 
extremely interesting.
One example is the reaction $p \bar{p} \to W^{\pm} b \bar{b} + anything,$ 
with  the two subprocesses $p \bar{p} \to W^{\pm} H^0$
($H^0 \to b \bar{b}$) and $p \bar{p} \to t b$ ($t \to W b$)
\cite{boos0}, which - 
together with several other SM diagrams - contribute
to the $W$\bb \,final state.
Another example is the reaction \cite{boos1}.

\begin{equation}
\gamma \quad e \longrightarrow \nu \quad b \quad \bar{b} \quad W^-
\label{eq:main}
\end{equation}

\begin{figure}[hbtp]
\begin{center}
\mbox{\epsfxsize=17cm\epsfysize=19cm\epsffile{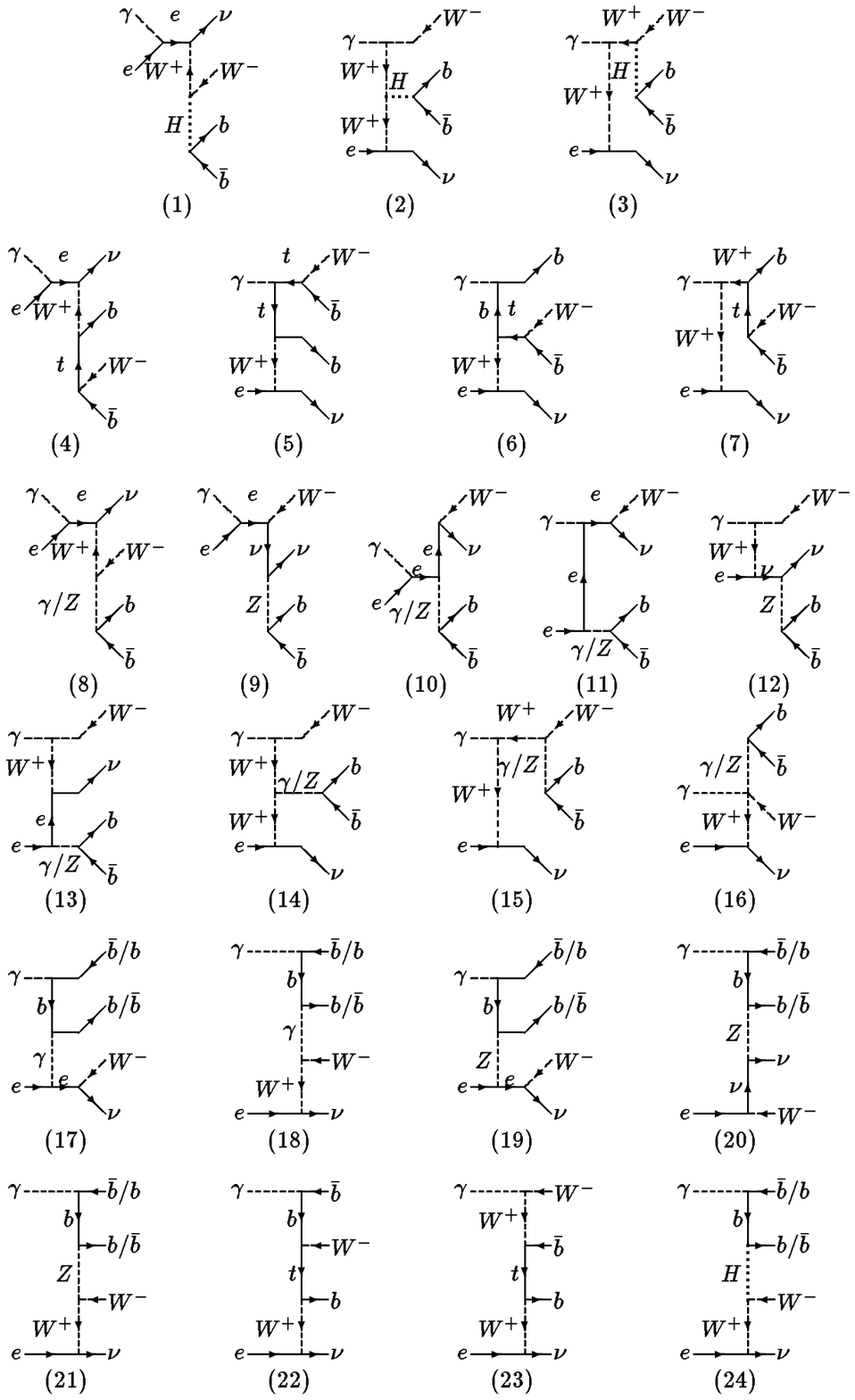}}   
\end{center}
\caption{ Feynman diagrams for the reaction \geinto \nbbw.}
\label{fig:feyn_ebbw}
\end{figure}

 Here, three out of 24 SM diagrams, shown in Fig. \ref{fig:feyn_ebbw}, 
involve associated Higgs boson production, 

\begin{equation}
\gamma \quad e \longrightarrow \nu \quad W^- \quad H^0,
\label{eq:1}
\end{equation}

\noi and four diagrams represent single top quark production,

\begin{equation}
\gamma \quad e \longrightarrow \nu \quad \bar{t} \quad b,
\label{eq:2}
\end{equation}

\noi with subsequent decays of the Higgs boson  (\hnull \into \bb)
and the top quark ($t$ \into $W b$). 

% feynman diagrams for eg ---> ebbw
%------------------------------------------------------------

Both reactions have been studied in the past \cite{boos2,jikia}
and their abundant rates were emphasized.
The associated Higgs production reaction (\ref{eq:1})
has a large sensitivity for probing 
anomalous $WWH$ coupling structures \cite{boos1}, 
whereas the single top reaction (\ref{eq:2}) 
is a unique tool for measuring the \vtb \ matrix element with 
very high precision \cite{boos1, jikia, boos3}.  

In this study we present results of complete
tree-level calculations of the
reaction \geinto \nbbw including in a model independent way
possible anomalous $Wtb$ and \hww couplings. 
Decays of unstable particles with correct spin structures
and contributions from all nonresonant diagrams
are taken into account.
The subreactions (\ref{eq:1}) and 
(\ref{eq:2}) are involved in the 2-to-4 body calculations
and, as will be shown, they can easily be extracted from
the inherent background leading to the same final state.
In our discussion we will follow
basically the papers \cite{boos1,boos4}.

The results presented have been obtained by means of the computer
package \CO \,3.2 \cite{comphep}.
As seen from the diagrams in Fig. \ref{fig:feyn_ebbw}, a number of
singularities exists in the s- and t-channels.
In the phase space integration by the adaptive Monte Carlo method, a
proper treatment of such singular behaviour is necessary in order to
obtain stable results.       
Singularities are smoothed by appropriate transformations of
variables, and the procedure of ref. \cite{pukhov} 
has been adopted in our calculations.

\begin{figure}[htbp]
\begin{center}
   \mbox{\epsfxsize=17cm\epsfysize=9cm\epsffile{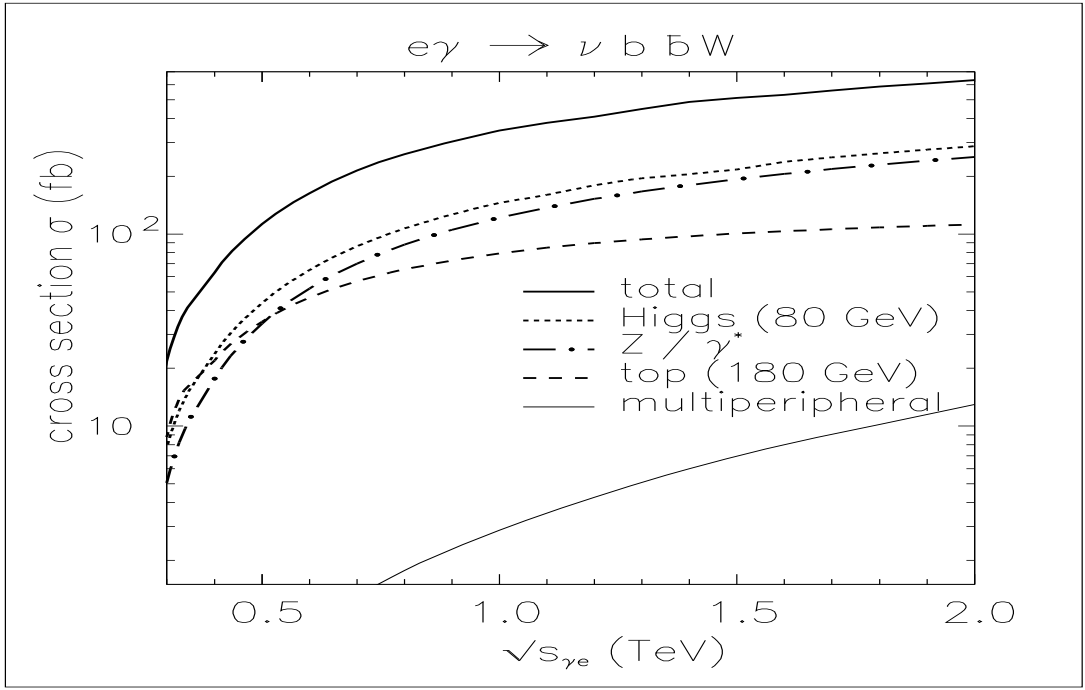}}
\end{center}
\caption{Total cross section for the reaction   \geinto \nbbw
as function of the $\gamma e$ cm energy.Also the individual
contributions of the Higgs, the top, the $Z/\gamma^*$ and the
multiperipheral diagrams are shown.}
\label{fig:2_nbbw}
\end{figure}

The total cross section for the reaction  
(\ref{eq:main}), \geinto \nbbw, and the  
subprocess contributions are shown in
Fig. \ref{fig:2_nbbw} as a function of \SQRTSGE 
for a Higgs
mass of \mh = 80 GeV and a top mass of \mt = 180 GeV.
% figure 2
%------------------------------------------------------------
All cross section rise with 
increasing energy over the whole energy range considered due to
the peresence of the t-channel diagrams.
If the \SQRTSGE \, is varried in the interval 0.5 - 2.0 TeV,
the total cross section and the cross sections for Higgs and top 
production vary  from 110 fb to 630 fb (total), 
from 45 fb to 290 fb (Higgs, \mh=80 GeV) and
from 35 fb to 110 fb (Top, \mt=180 GeV) respectively, and 
they are large enough to be very interesting for more 
detailed studies.

However in order to obtain 'realistic' estimations of event
rates, the reaction cross   
sections have to be convoluted with some backscattered
photon flux.
We have, as an example, adopted in our calculations the widely used 
photon spectrum (see the paper \cite{ginzburg}).
The convolution leads to a decrease of the cross section by a factor
of 2.5 - 1.5 depending on the collider energy; the rate reduction
is larger at smaller energies. After the convolution
the total cross section of reaction (\ref{eq:main}) varies from 40 fb to 
420 fb, the Higgs cross section  
changes from 16 fb to 185 fb for $M_H = 80$ GeV (8 fb to 160 fb 
for $M_H = 140$) and the top contribution from 14 fb to 90 fb 
between \SQRTSGE \, = 0.5 - 2.0 TeV. It is encouraging that
even after such a degradation of the basic cross sections, 
an electron-photon collider can considerably improve the
physical capabilities of Higgs and top studies.

%=============================================================================
\section{$W t b$ coupling}
%\label{sec5}
%=============================================================================

In this section we consider the precision of the measurement for the \vtb \ 
matrix element and a probe of a possible anomalous $Wtb$ coupling
in a model independent way.

Measurements of \vtb \, or the partial width $\Gamma_{tWb}$, which are
related in the SM, are known to be nontrivial.
The study of the single top quark reaction 
\eeinto \entb \, at high energies \cite{boos3} offers the 
possibility to
obtain a relatively precise value of \vtb.
In this channel, the \vtb \, measurement capability relies mainly on
the Weizs\"{a}cker-Williams photon exchange contributions, $\gamma^*
e$ \into $\nu b t$.
Using however
the laser backscattered high energy photon beam instead of $\gamma^*$ the
cross sections are typically enhanced by a factor of 3 to 5.

\begin{figure}[htbp]
\begin{center}
   \mbox{\epsfxsize=17cm\epsfysize=9cm\epsffile{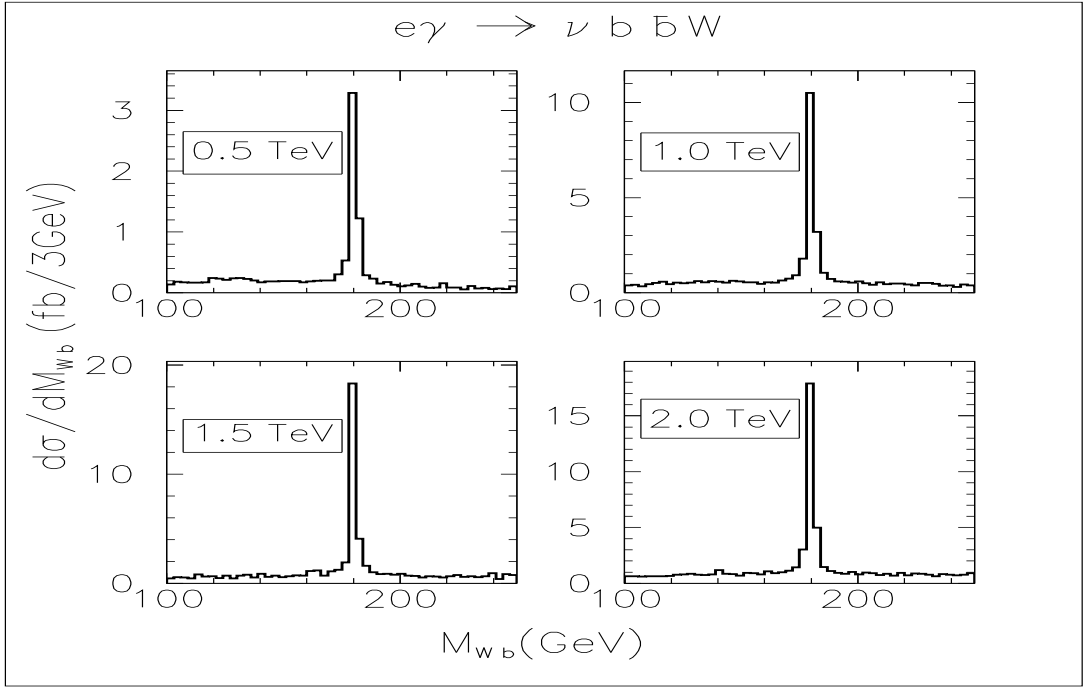}}
\end{center}
\caption{Differential cross sections d$\sigma$/d$M_{Wb}$
  of reaction (\ref{eq:main}) as
  function of $M_{Wb}$.}
\label{fig:4}
\end{figure}

The single top production rate is directly proportional to
\vtb$^2$. However, the top quark is not observed directly; it decays 
into a $W$ and a $b$ quark, leading to the final state \nbbw. There are,
as mentioned, several other contributions to the same final state
which are not proportional to \vtb. Fortunately, 
the extraction of the top out of this 4-body final
state can be easily achieved with a small lost of the single top rate
as seen from Fig. \ref{fig:4}, where the invariant masses
of the $W$ and the $b$ are shown for four energies.   
Selection of top events requires only a cut around \mt \, and no
further demands.
 
% figure 4
%------------------------------------------------------------

Using expected \ee \, luminosities as proposed in ref. \cite{wiik}, an
$e-\gamma$ conversion factor of 0.8 and a 30\% $\nu b t$ event 
detection probability (due to cuts to observe the top decay products
and the $b$-jet and to eliminate backgrounds; major backgrounds are
expected from the reactions \geinto $\nu W Z$ and \geinto $eW^+W^-$),
the two-standard deviation errors on \vtb \,
are shown in Tab. \ref{tab:7}.
\begin{table}[htbp]\centering
\caption{Two-standard deviation error of \vtb \, expected for the
  annual luminosities as indicated. }
\begin{tabular}{lllll}
                                &      &      &      &       \\
\hline\noalign{\smallskip}
                                &      &      &      &       \\
\SQRTSEE, TeV                    &  0.5 & 1.0  & 1.5  & 2.0   \\
                                &      &      &      &       \\
\hline\noalign{\smallskip}
\hspace{3cm} & \hspace{2cm} & \hspace{2cm} & \hspace{2cm} & \hspace{2cm} \\
 $\cal{L}$ fb$^{-1}$        &  50  & 200 & 300   & 500   \\
 & &  &  & \\
 $\delta$\vtb        &  8\% & 2\% & 1.5\% & 1\% \\
 & &  &  & \\
\hline\noalign{\smallskip}
\end{tabular}
\label{tab:7}
\end{table}
As can be seen, the CKM matrix element \vtb \, can be probed with high
accuracy. Our error obtained at \SQRTSEE of order of 
0.5 TeV is similar to the 
one-standard deviation error expected at the upgraded Tevatron 
and LHC \cite{boos-willenbrock}.
However, a higher energy $\gamma e$ collider provides a better 
determination of \vtb \, .

In order to probe an anomalous $Wtb$ coupling 
in a model independent way, we use the effective 
Lagrangian approach \cite{buchmueller,gounaris1} with 
notations in the unitary gauge as given in ref. \cite{kane}.
The Lagrangian ${\cal L}$  contains 
only necessary vertices for the process (\ref{eq:2}):

\begin{eqnarray}
{\cal L} & \left.=  \frac{g}{\sqrt{2}}\right[ & 
 W_{\nu}^-\bar{b}(\gamma_{\mu}F^L_1P_- + F_1^RP_+) t \nonumber \\ 
 & & - \left.\frac{1}{2M_W} W_{\mu\nu}
\bar{b}\sigma^{\mu\nu}(F_2^LP_- + F_2^RP_+) t  \right] + {\rm h.c.} 
\label{eq:lagrangian_anom}
\end{eqnarray}

\noi with $W_{\mu\nu} = 
D_{\mu}W_{\nu} - D_{\nu}W_{\mu}, D_{\mu} = \partial_{\mu} - i e A_{\mu}, 
P_{\pm} = 1/2(1 \pm \gamma_5)$ and $\sigma^{\mu\nu} 
= i/2(\gamma_{\mu}\gamma_{\nu} - \gamma_{\nu}\gamma_{\mu})$.
The similarity of the  $\sigma^{\mu\nu}$-connected operators with the 
QED anomalous magnetic moments prompts the name 
`magnetic type' for the operators and their associated vertices.  
Within the Standard Model,  
\lone = \vtb \,and \rone = $F_2^{L,R} = 0$.
Terms containing  $\partial_{\mu} W^{\mu}$ 
are omitted in the Lagrangian.
They can be recovered by applying  the quantum
equation of motion through operators of the original Lagrangian
\cite{gounaris1}.
We assume CP conservation with  $F_i^{L,R} = F_i^{*L,R}$.
The corresponding Feynman rules in the unitary gauge, obtained from the 
effective Lagrangian, are listed in the Appendix of \cite{boos4}.
These rules for the new vertices have been implemented into 
the program \CO \,3.2.

\begin{figure*}[h!b]
\begin{center}
\mbox{\epsfxsize=17cm\epsfysize=7cm\epsffile{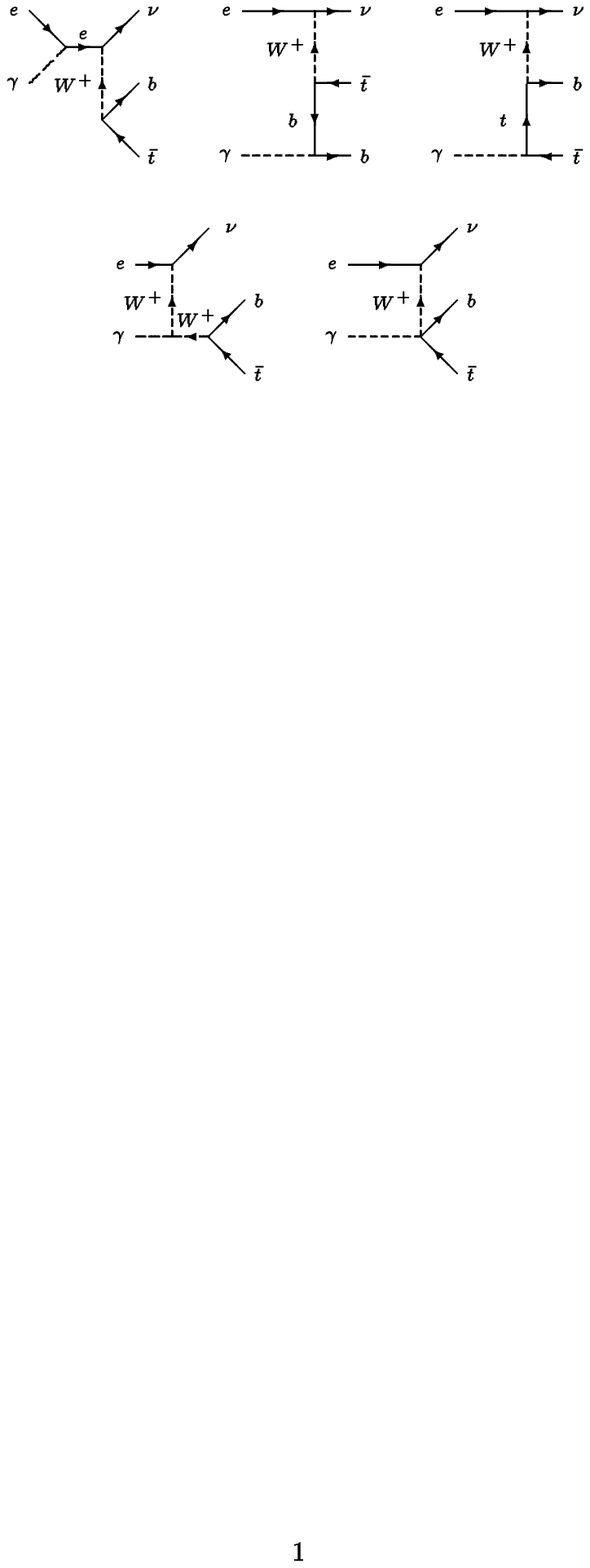}}
%\picplace{19cm}
\end{center}
\caption{ Feynman diagrams for the reaction \gentb.}
\label{fig:feyn}
\end{figure*}

The tree-level diagrams contributing to the subreaction  \gentb
\,are shown in Fig. \ref{fig:feyn} where the last non-SM diagram
with the four-point $\gamma W t b$ vertex is needed in order to ensure  
${\cal U}(1)$ gauge invariance. The new vertex $\gamma W t b$
contains  the sole contribution from
the magnetic type of operators proportional to $F_i^{*L,R}$ and follows from 
the effective Lagrangian.

% feynman diagrams for eg ---> nbt
%------------------------------------------------------------

\begin{figure*}[hbtp]
\begin{center}
\mbox{\epsfig{figure=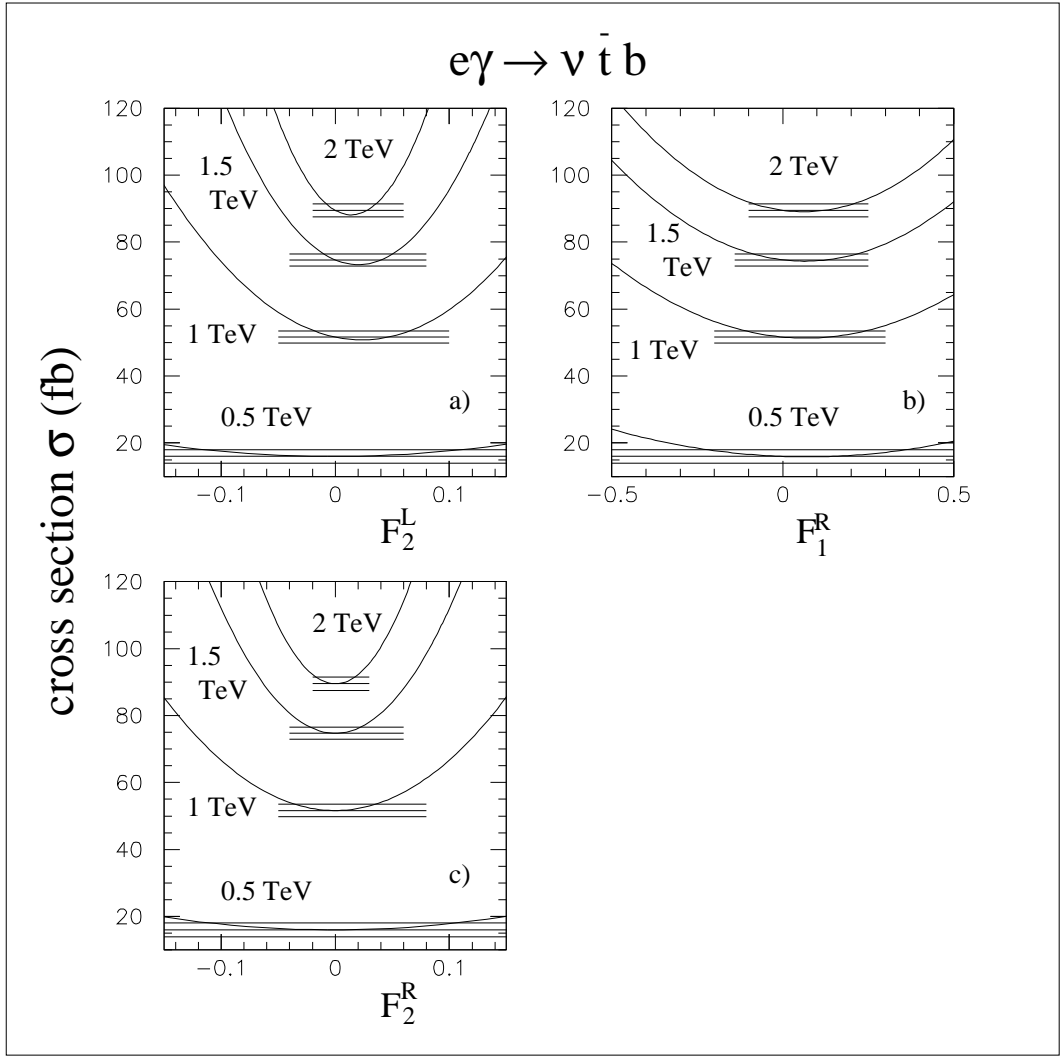,width=14cm}}
%\picplace{19cm}
\end{center}
\caption{Cross sections of the reaction \gentb \, as functions of the
  anomalous couplings \ltwo, \rone
  \,and \rtwo at \SQRTSEE = 0.5, 1.0, 1.5 and 2.0 TeV. The horizontal 
  lines show the SM  values with the two standard deviation errors expected.}
\label{fig:2}
\end{figure*}

Fig.\ref{fig:2} shows the variation of the single top cross 
section as function  of the  
anomalous couplings  \rone, \ltwo, \rtwo at fixed SM value for
\vtb , at four 
cm energies \SQRTSEE = 0.5, 1.0, 1.5 and 2.0 TeV.
% f variation 
%------------------------------------------------------------

Each of the figures 5a-c reflects a possible deviation of the different
anomalous couplings around zero with the other F-parameters fixed to
the SM-values.
A common feature is an increasing sensitivity with growing energies
and an enhancement of the cross section when the couplings deviate
from the SM value.
As expected from the additional power of momentum 
the  $F_2^{L,R}$ couplings represent a
much higher sensitivity to variations from the SM than the \rone. 

With the same luminosities \cite{wiik} and the   
event detection efficiency 
of 30\% as above we calculated limits of the variation 
of \rone, \ltwo and \rtwo \,within 
two standard deviations of the SM cross section.
As can be seen from Table \ref{tab:1}, the limits of the anomalous couplings 
obtained are in the interesting region \cite{peccei} of 
\begin{equation}
\frac{\sqrt{m_b m_t}}{v} \sim 0.1
\end{equation}
and do not exceed the unitary violation bounds \cite{gounaris2}
 in the one TeV scale of 
\begin{equation}
F_2^{R,L} \sim 0.8 \hspace{.5cm}{\rm and} \hspace{.5cm}  F_1^R \sim 0.6.
\end{equation}
\begin{table*}[htbp]\centering
\caption{Limits for the anomalous 
couplings $F_i^{L,R}$ obtained from the two 
standard devitation critera as described in the text  
   for annual luminosities as indicated.}
\begin{tabular}{lllll}    
                                &      &      &      &       \\
\hline\noalign{\smallskip}
                                &      &      &      &       \\
\SQRTSEE, TeV                    &  0.5 & 1.0  & 1.5 & 2.0   \\
                                &      &      &      &       \\
\hline\noalign{\smallskip}
\hspace{3cm} & \hspace{2cm} & \hspace{2cm} & \hspace{2cm} & \hspace{2cm} \\
 $\cal{L},$ fb$^{-1}$        &  50  & 200 & 300   & 500   \\
 & &  &  & \\
\hline\noalign{\smallskip}
 & &  &  & \\
 $\delta$\ltwo         &  -.1/.1 & -.020/.065 & -.01/.05 & -.008/.035 \\

 $\delta$\rtwo         &  -.1/.1 & -.035/.035 & -.022/.022 & -.016/.016 \\

 $\delta$\rone         &  -.20/.35 & -.12/.25 & -.09/.22 & -.08/.20 \\

 & &  &  & \\
\hline\noalign{\smallskip}
\end{tabular}
\label{tab:1}
\end{table*}
For comparison, recent studies of 
single top production rates including anomalous couplings  at the Tevatron 
indicate the  
bounds   -0.5 $\less F_1^R \less$ 0.5 
\cite{c-p, boos5},  
  -0.1 $\less F_2^L \less$ 0.2 and -0.2 $\less F_2^R \less$ 0.2 
\cite{boospp} which are comparable with our results  expected at NLC
energies of 0.5 TeV.
At energies above 0.5 TeV we obtain significantly higher 
sensitivities (see Table \ref{tab:1}).

\begin{figure*}[h!b]
\begin{center}
\mbox{\epsfig{file=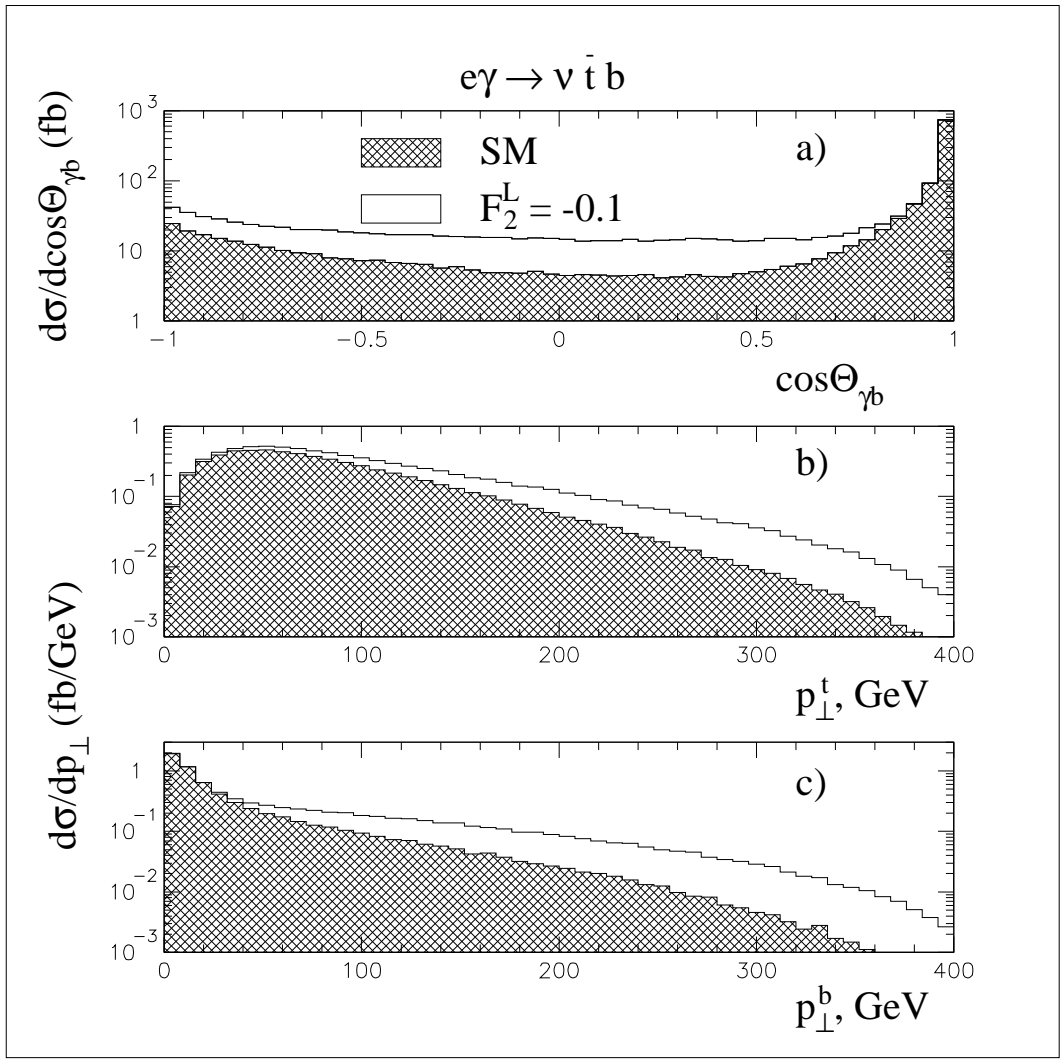,width=14cm}}
%\picplace{19cm}
\end{center}
\caption{The  cos$\Theta_{\gamma b}$, 
  \Ptt \,and \Ptb distributions at 1.0 TeV. Compared are the SM
  predictions (hatched areas) with expectations from an 
  anomalous coupling \ltwo = -0.1.}
\label{fig:11}
\end{figure*}

The existence of anomalous couplings also affects  
production properties of the final state particles of reaction (\ref{eq:2}).
As an example, Fig.\ref{fig:11}a-c show the differential cross
sections d$\sigma/$dcos$\Theta_{\gamma b}$, 
d$\sigma$/d\Ptt and  d$\sigma$/d\Ptb  
 expected for  \ltwo  = -0.1, \rtwo = \rone = 0 and \lone = SM 
value (open areas), compared with the SM predictions 
(hatched areas)\footnote{The angle $\Theta_{\gamma b}$ 
is defined as the angle of the $b$-quark with respect 
to the incident photon  direction in the \ee \,rest frame.}. 
% cos theta, pt distributions
%------------------------------------------------------------
In particular, the SM angular distribution 
d$\sigma/{\rm d}\cos \Theta_{\gamma b}$ in 
Fig. \ref{fig:11}a has a broad minimum around  $\cos \Theta_{\gamma b}
\sim$ 0 - 0.5. 
This behavior is due to the existence of the so called radiation zero
of the 2-to-2  body process \qq \into $W\gamma$   \cite{samuel}
and its time-reversed reaction 
 $\gamma W \rightarrow \bar{t} b$
 as the most
important subreaction for our consideration.
In our case, the incident $\gamma$ spectrum and the off-shell
character of the $W$-boson in addition to the contribution of the
first diagram of Fig. 1 washed out this zero to a broad minimum.
Nevertheless, for anomalous coupling contributions the minimum becomes 
significantly higher.

In the Lagrangian (4), the (V+A) operator which is 
proportional to the \rone \,coup\-ling has only an overall numerical factor
and leads to a simple shift of the  \Pt \,distributions.
On the other hand,
as was mentioned the new  anomalous magnetic type vertices 
contain an additional 
power of momentum (see \cite{boos4}) and therefore the transverse momentum 
distributions of the 
$t-$ and $b$-quark deviate from the SM  expectations.
As a consequence, 
such different behaviour allows to separate contributions
of the (V+A) operator from the magnetic type ones.  
Fig. \ref{fig:11}b and c show
an  excess at high \Pt for both, the  \Ptb \,and the \Ptt \,distributions. 
Clearly, cuts in the tranverse momenta and angular distributions
should lead to significantly more stringent constraints in $F_i^{R,L}$.
For illustration purpose, we require
\Ptb $>$ 40 GeV, \Ptt $>$ 80 GeV and $\Theta_{\gamma b}$ $> 10^0$ 
for the cross section calculation at 1 TeV.
The bounds are improved to -0.012 $<$ \ltwo $<$ 0.058 which 
should be compared with 
 -0.020 $<$ \ltwo $<$ 0.065 (see Tab. \ref{tab:1}).

A further possibility for studying anomalous couplings 
could be the measurement of 
the top quark partial decay width \cite{gounaris2} described by
the same effective Lagrangian ${\cal L}$. 
Here, the partial decay width is extracted from the 
single top production rate and therefore the measurement 
is not independent from the procedure given above.

An unpolarized $tt$-pair production measurement at 
hadron collisions would only deliver 
the branching ratio of the top quark 
into $W$-boson and $b$-quark comparing the single and 
double $b$-tagging rates \cite{dan}.
Calculations show that the branching ratio is very insensitive to
variations of the F-parameters.
Even for extreme values of the parameters in the range of $\pm$1 the
branching fraction varies from 99.7\% to 99.9\%. 
Since the precision of the 
determination of the branching ratio is of the order of 10\%, a 
deviation from the SM value of 99.8\% due to the influence of 
anomalous couplings  will not be visible.

%=============================================================================
\section{Probing the \hww coupling}
%\label{sec6}
%===========================================================================
Reaction (\ref{eq:main}), \geinto \nbbw, involves also
significant Higgs production
with a rate directly proportional to the \hww coupling.
In the SM the Higgs-vector boson vertices are uniquely determined.
In the following we parametrize possible non-SM $WWH$ coupling 
by introducing an effective non-renormalizable Lagrangian which
preserves the SM gauge group
\begin{equation}
{\cal L}_{eff} = {\cal L}_{SM} +
\sum_{k=1}^{\infty}\frac{1}{(\Lambda^2)^k} \sum_i
f_i^{(k)}Q_i^{d_k} \hspace{1cm} ,
\label{eq:lagrangian}
\end{equation}
where $d_k = 2 k + 4$ denotes the dimension of operators and
$\Lambda$ is the energy scale of new interactions.
We limit ourselfs to
the complete set of the effective dimension-6 operators as
outlined in ref. \cite{buchmueller}.
Under this restriction phenomenological
applications to anomalous Higgs couplings have been discussed in
\cite{gounaris1}-\cite{kilian}.
One can show that from the four operators involving the $WWH$ coupling
there are only two \cite{kilian} which preserve the 
custodial SU(2) symmetry \cite{custodial}
relevant for the Higgs subprocess \geinto $\nu W H$ :
\begin{equation}
\frac{1}{\Lambda^2}\left\{\frac{1}{2}f_{\varphi}\partial_{\mu}  
(\Phi^+\Phi)\partial^{\mu}(\Phi^+\Phi)
+ f_{WW}\Phi^+({\hat W}_{\mu\nu}\hat W^{\mu\nu})\Phi\right\}
\hspace{1cm} .
\label{eq:1overlambda}
\end{equation}
Such an effective \hww interaction has been also implemented in the program 
\CO \, (see the Appendix in \cite{boos1}) with the following
definition of the parameters $F_i$ :
\begin{equation}
F_{\varphi}/(1TeV^2) = f_{\varphi}/\Lambda^2 \hspace{1cm}   \hspace{1cm}
F_{WW}/(1TeV^2) = f_{WW}/\Lambda^2
\label{eq:F}
\end{equation}

\begin{figure}[htbp]
\begin{center}
   \mbox{\epsfxsize=17cm\epsfysize=9cm\epsffile{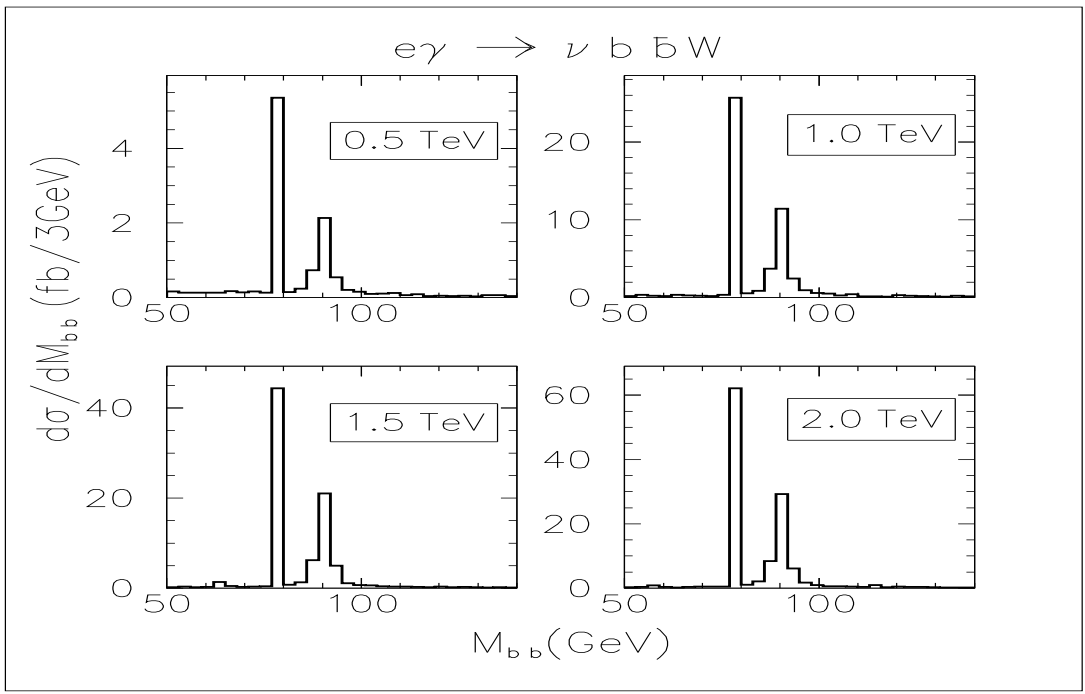}}
\end{center}
\caption{Differential cross sections as a function of the \bb \,
  invariant mass at \ee \, cm energies of 0.5, 1.0, 1.5 and 2.0 TeV.
  Clear \hnull
  \, and $Z$ peaks are visible on a very small background.} 
\label{fig:fmassb}
\end{figure}

\begin{figure}[htbp]
\begin{center}
   \mbox{\epsfxsize=17cm\epsfysize=9cm\epsffile{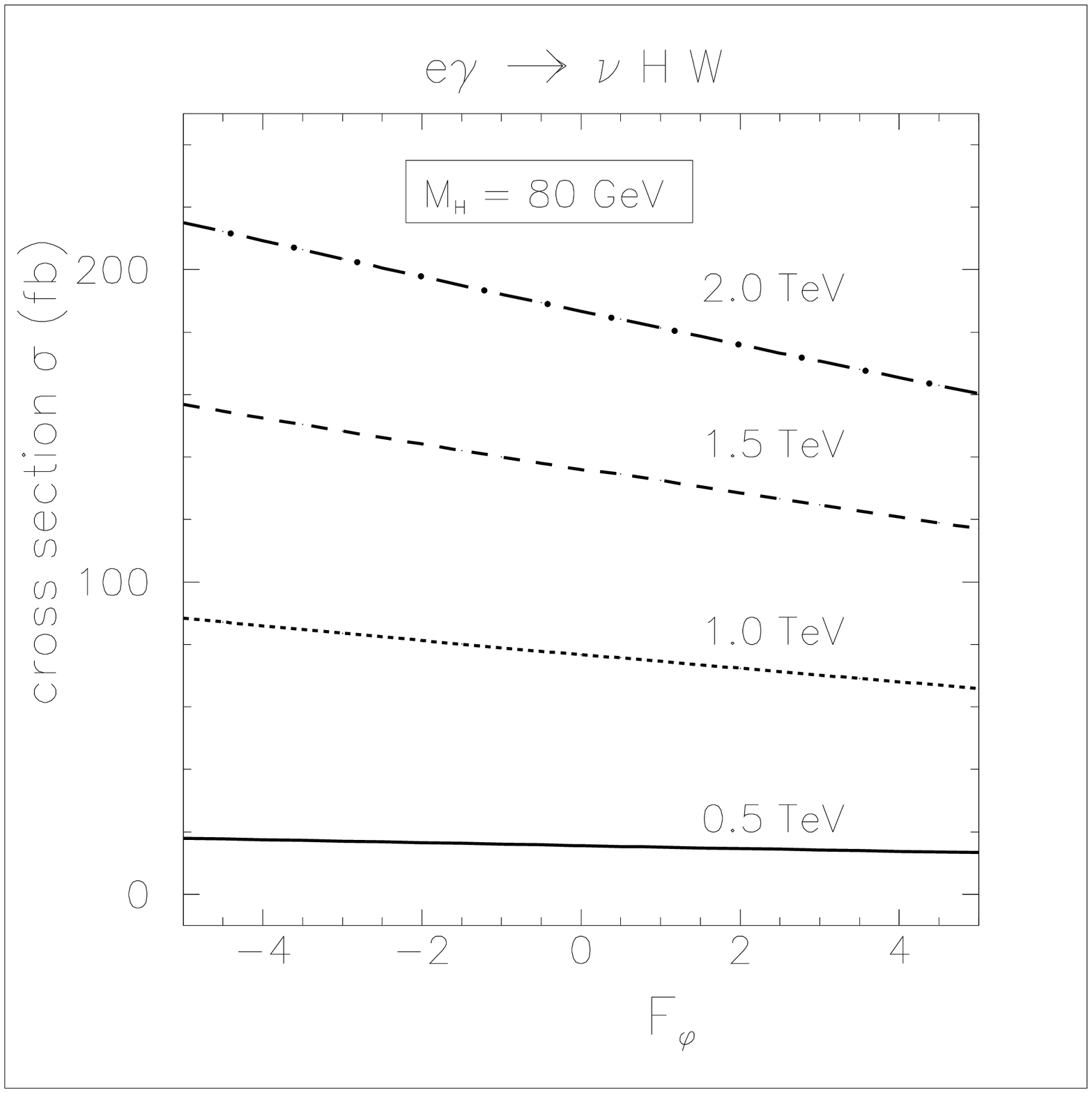}}
\end{center}
\caption{Higgs cross sections  as function
of the parameter $F_{\varphi}$ with 
  $F_{WW}$ = 0, at \ee \, cm energies  of
  0.5, 1.0, 1.5 and 2.0 TeV for \mh =
  80 GeV.}  
\label{fig:5}
\end{figure}

\begin{figure}[htbp]
\begin{center}
   \mbox{\epsfxsize=17cm\epsfysize=9cm\epsffile{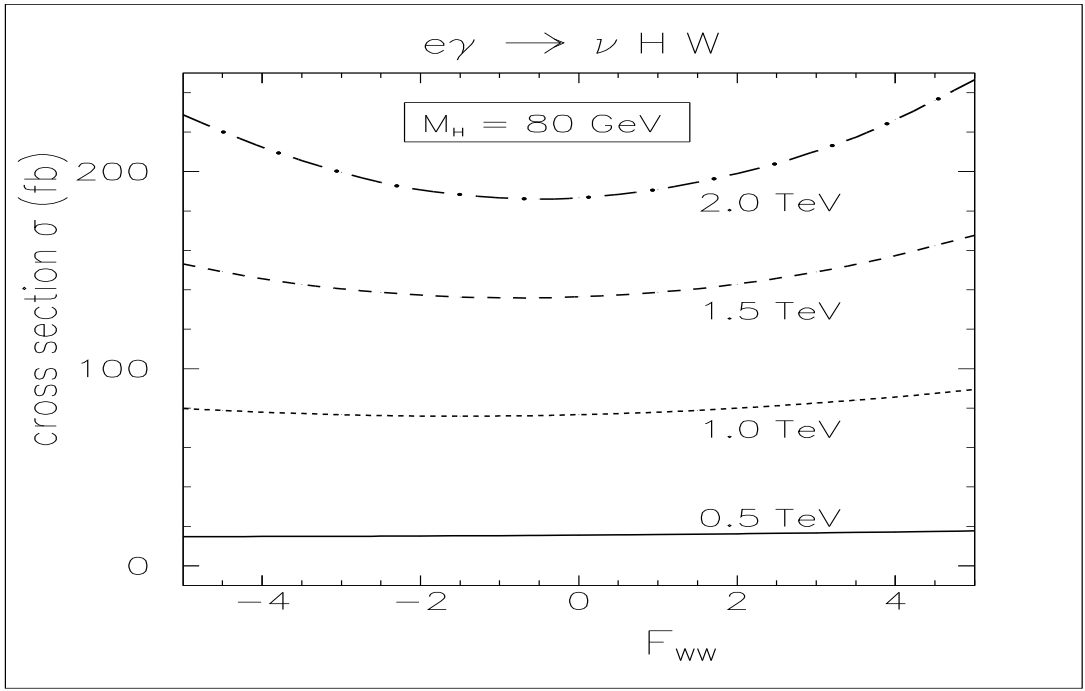}}
\end{center}
\caption{Higgs cross sections as functions
of the parameter $F_{WW}$ with
  $F_{\varphi}$ = 0, at \ee \, cm energies of 0.5, 1.0, 1.5 and 2.0 TeV 
for \mh
  = 80 GeV.}
\label{fig:6}
\end{figure}

Probing the \hww \, coupling involves calculating the
dependence of the cross section of the subreaction
\geinto $\nu W H$ on the
parameters $F_i$ and comparison with the SM expectation.
As in the case of the single top production, 
the $\nu W H$ events also can be extracted 
from the 4-body final state 
$\nu b\bar{b}W$ of reaction 
(\ref{eq:main}) by imposing a cut on the
\bb \, invariant mass, \mh - 3 GeV $< M(b\bar{b}) <$ \mh + 3 GeV,
as seen in Fig. \ref{fig:fmassb}.
% figure mbb
%------------------------------------------------------------
Fig. \ref{fig:5}(\ref{fig:6}) shows the Higgs cross section as
function of $F_{\varphi}(F_{WW})$ for $F_{WW}(F_{\varphi})$ = 0, at \SQRTSEE
= 0.5, 1.0, 1.5 and 2.0 TeV for \mh = 80 GeV.
% figure 5  
%------------------------------------------------------------   
% figure 6
%------------------------------------------------------------
In order to estimate the ranges of $F_i$ which can
be probed within our assumptions, we determined those variations of
the $F_i$ which leave the cross section unchanged within 2 s.d.
from the SM value.
Only statistical errors of the cross sections were
considered taking into account once more the integrated luminosities of
Tab. \ref{tab:7}, an $e-\gamma$ conversion factor of 0.8 and a 30\%
$\nu H W$ detection probability.
The intervals of $F_i$ obtained are presented in
Tab. \ref{tab:8}.
\begin{table}[htbp]\centering
\caption{Range of   $|F_{\varphi}|$ and  $|F_{WW}|$ obtained from the
  two-standard deviation criteria as described
  in the text.}

\begin{tabular}{lllll}
                                &      &      &      &       \\
\hline\noalign{\smallskip}
\hline\noalign{\smallskip}
                                &      &      &      &       \\
\SQRTSEE, TeV                    &  0.5 & 1.0  & 1.5  & 2.0   \\
                                &      &      &      &       \\
\hline\noalign{\smallskip}
\hspace{3cm} & \hspace{2cm} & \hspace{2cm} & \hspace{2cm} & \hspace{2cm} \\
 $|F_{\varphi}|$                   &  5.0  & 1.0  & 0.6   & 0.4   \\
                                &      &     &       &       \\
 $|F_{WW}|$                     &  9.0  & 2.5 & 2.0    & 1.0    \\
                                &      &     &       &       \\   
\hline\noalign{\smallskip}
\hspace{3cm} & \hspace{2cm} & \hspace{2cm} & \hspace{2cm} & \hspace{2cm} \\
 $|F_{\varphi}|$                   &  5.0  & 1.0  & 0.6   & 0.4   \\
                                &      &     &       &       \\
 $|F_{WW}|$                     &  9.0  & 2.5 & 2.0    & 1.0    \\
                                &      &     &       &       \\
\hline\noalign{\smallskip}
\end{tabular}
\label{tab:8}
\end{table}

Clearly, the larger the cm energy the more sensitive the cross section
becomes to a modification of the \hww \, coupling.
The analyses of the two-body reactions $e^+e^- \rightarrow H \gamma$ or $HZ$
\cite{gounaris2, kilian} revealed higher sensitivities for these operators
at energies below 1 TeV. At the energies 1-2 TeV the reaction 
\geinto $\nu W H$ becomes comparable or slitely more sensitive. 

%=============================================================================
\section{Summary}
%\label{sec7}

Results \cite{boos1, boos4} of 
the complete tree-level calculation of the 
reaction \geinto \nbbw \, at cm energies  0.5 to 2.0 TeV are presented 
and discussed.
The reaction is very interesting on its own because it involves  
at the same time single top production, \geinto $\nu \bar{b} t$, and
associated Higgs production, \geinto $\nu H W$, with subsequent decays
of $t$ \into $W b$ and $H$ \into \bb, respectively.
Therefore, both three-body subreactions already studied in previous
publications are analyzed in an extended manner taking into account
interferences between different subchannels, the
irreducible background, and contributions from the anomalous
$Wtb$ and \hww \, couplings. It is 
demonstrated that
both subreactions can be easily extracted fron the 4-body final state.

The event rate for the reaction \geinto $\nu t b$, which is large even after
folding with an energy spectrum of the backscattered photon beam
and making reasonable assumptions on collider
luminosities and detection probabilities,
provides the best sensitive 
measurement, compared to other collision processes,  for the CKM 
matrix element \vtb as well as for probing the
anomalous $Wtb$ couplings in a model independent way.

The reaction \geinto $\nu H W$ allows to probe the \hww \, coupling
and to measure parameters of dimension-6 operators in the effective 
Lagrangian.
It has been found that at 0.5 TeV the accuracy obtained on these
parameters is not sufficient to make this measurement
sensitive to new physics while at energies \SQRTSEE = 1-2 TeV
the \hww \, coupling can be probed with high sensitivity and
deviations from the Standard Model could show up.

%--------------------------------------------------------------------------
%     acknowledgments
%--------------------------------------------------------------------------

\section*{Acknowledgments}
The work has been supported in part by the 
RFBR grant 96-02-19773a, and by the grant 95-0-6.4-38 of 
the Center for Natural Sciences of State Committee for Higher Education
in Russia. E.B. is grateful to the Deutsche 
Forschungsgemeinschaft (DFG) for the financial support.

%--------------------------------------------------------------------------
%       bibliography
%--------------------------------------------------------------------------

\end{document}